\documentclass[]{spie}
\usepackage[]{graphicx}
\usepackage{aas_macros}

\title{Kernel-phases for high-contrast detection beyond the resolution
  limit}

\author{Frantz Martinache\supit{a}
\skiplinehalf
\supit{a}Subaru Telescope, 650 N. A'ohoku Place, HI 96720, USA \\
}

\authorinfo{email: frantz@naoj.org}

\begin{document}
\maketitle

\begin{abstract}
The detection of high contrast companions at small angular separation
appears feasible in conventional direct images using the
self-calibration properties of interferometric observable
quantities. In the high-Strehl regime, available from space borne
observatories and using AO in the mid-infrared, quantities comparable
to the closure-phase that are used with great success in non-redundant
masking inteferometry, can be extracted from direct images, even taken
with a redundant aperture. These new phase-noise immune observable
quantities, called Kernel-phases, are determined a-priori from the
knowledge of the geometry of the pupil only. Re-analysis of HST/NICMOS
archive and other ground based AO images, using this new Kernel-phase
algorithm, demonstrates the power of the method, and its ability to
detect companions at the resolution limit and beyond.
%The detection of high contrast companions at small angular separation
%appears feasible in conventional direct images using the
%self-calibration properties of interferometric observable
%quantities. In the high-Strehl regime, soon to become widely available
%thanks to the coming generation of extreme Adaptive Optics systems on
%ground based telescopes, and already available from space, quantities
%comparable to the closure-phase that are used with great success in
%non-redundant masking inteferometry, can be extracted from direct
%images, even taken with a redundant aperture. These new phase-noise
%immune observable quantities, called Kernel-phases, are determined
%a-priori from the knowledge of the geometry of the pupil
%only. Re-analysis of HST/NICMOS archive and other ground based AO
%images, using this new Kernel-phase algorithm, demonstrates the power
%of the method, and its ability to detect companions at the resolution
%limit and beyond.
\end{abstract}

\section{Introduction}

Imaging in the high angular resolution regime comes down to solving
the problem of deconvolving an unknown object function from an ever
changing point spread function (PSF). The development of adaptive
optics (AO), now ubiquitous on major ground based observatories,
dramatically changed this arena, by turning seeing limited images into
diffraction limited ones.

High contrast detection is however currently limited by residual
aberrations, responsible for the presence of speckles in the image.
Different schemes have been developed to sort out the PSF from the
object function in images, and a very successful approach is to use
some form of diversity in the PSF: angular differential imaging
\cite{2006ApJ...641..556M} for instance, uses field-rotation to
differentiate true companions from static speckles. Aggressive
alternatives are becoming available with updated AO systems, often
refered to as extreme-AO systems, with active optics that can be used
to introduce phase diversities, for instance creating speckle free
regions in the image \cite{1995PASP..107..386M} or estimating the
coherence of speckles \cite{2010PASP..122...71G}.

This paper presents an alternative approach, based on an
interferometric point of view of image formation. The approach builds
on the recent development of non-redundant aperture masking
interferometry \cite{1988AJ.....95.1278R, 1989AJ.....97.1510N}. 
The technique, first used in 1873 by Stephan in an attempt to measure
the angular diameter of stars with the 80-cm baseline of the telescope
of the Observatoire de Marseille, was revived by speckle
interferometry \cite{1970A&A.....6...85L}, the availability of large
telescopes \cite{2000PASP..112..555T}, and the generalization of AO
\cite{2006SPIE.6272E.103T}.

Non-redundant masking interferometry takes advantage of the
self-calibrating properties of an observable quantity called
closure-phase \cite{1958MNRAS.118..276J}. This remarkable quantity
exhibits a compelling property: it rejects all residual phase errors
on the interferometer pupil. Moreover, because it is determined from
the analysis of the final science detector, and not estimated from a
separate ``sensing'' channel, it is immune to non-common path errors
that are partly responsible for the presence of quasi-static speckles
in regular images.

Once extracted, the closure phase can then be used as input of a
parametric model, for instance of a binary star of variable
characteristics (angular separation, position angle and luminosity
contrast), to confirm or infirm the presence of a companion around a
given source, while uncertainties provide contrast detection limits.
This approach was for instance successfully used several references
\cite{2006ApJ...650L.131L, 2007ApJ...661..496M, 2008ApJ...678..463I,
  2008ApJ...679..762K, 2009ApJ...695.1183M},
that typically report sensitivity of 5-6 magnitudes in the
near-infrared at angular separations ranging from 0.5 to 4
$\lambda/D$.

It was demonstrated that the notion of closure phase, requiring a
strictly non-redundant aperture can be generalized to arbitrarily
shaped (i.e. including redundant) pupils, if the wavefront quality is
sufficient \cite{2010ApJ...724..464M}. 
This generalization of the closure-phase is coined Kernel-phase, since
these closure relations form a basis for the null-space (or Kernel) of
a linear operator. This paper first introduces the idea of generalized
closure-phases, formulating the problem in terms of linear
algebra. This new formulation is then used to show that it is not
restricted to non-redundant apertures, if the wavefront quality is
sufficient. Finally, the paper reports on a successful application of
the Kernel-phase approach to a set of ground-based data, opening the
doors of the super-resolution regime to a large number of observing
programs.

\section{Generalized closure-phase}

\begin{figure}
\centerline{\includegraphics[width=0.9\textwidth]
  {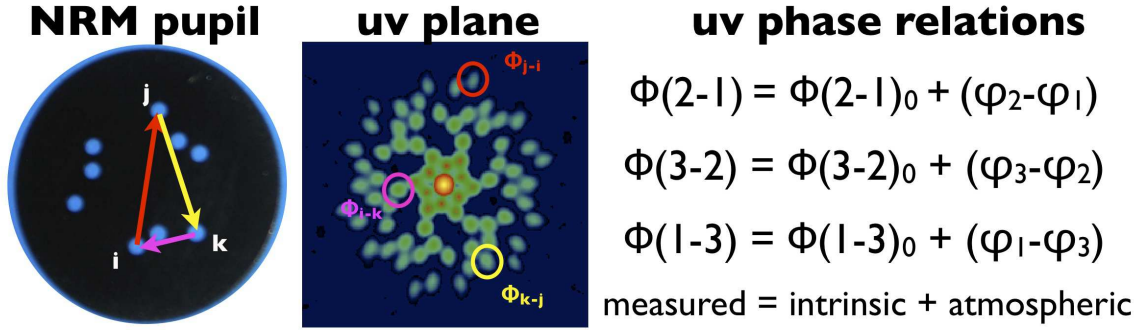}}
\caption{Example of closure phase relation. Superimposed on the image
  of the non-redundant pupil geometry shown in the left panel, is one
  of the possible closure triangles along which a closure relation can
  be formed.
  The central panel shows the powerspectrum of one image acquired with
  one such pupil: it is made of distinct regions, often refered to as
  splodges, each of which is associated to a baseline in the
  pupil. The three splodges associated to the baselines chosen in the
  left panel are highlighted.
  On the right, are written relations for the phases of each splodge:
  the measured phase is the sum of a term intrinsic to the target
  being observed (the ``true'' phase), and an atmospheric term: the
  piston along the baseline.
  The reader will quickly observe that by adding these three relations
  together, the atmospheric term simply vanishes, leading to a new
  observable quantity, called the closure-phase.
}
\label{fig:nrm}
\end{figure}

To understand how to generalize the idea of closure phase, it is
useful to go back to the non-redundant masking (NRM) scenario and
reformulate the problem in new terms.
Fig. \ref{fig:nrm} shows how the problem is traditionally presented. A
triangle of baselines in a non-redundant pupil is highlighted along
with a typical powerspectra. Expressions for the three phase terms
measured in the uv plane are listed.
By adding the three measured phases, the piston terms cancel out: the
resulting quantity, called a closure-phase, contains information about
the observed source only, robust to piston and wavefront errors,
making it extremely interesting.

The 9-hole NRM pupil presented in Fig. \ref{fig:nrm} allows to measure
36 distinct phases in the uv plane. These 36 relations can be written
together in a matrix form:

\begin{equation}
  \Phi = \Phi_0 + \mathbf{A} \cdot \varphi,
  \label{eq:transfer}
\end{equation}

\noindent
where $\Phi$ is a 36-component vector coding the uv plane phases,
$\Phi_0$ a vector of same size, coding the true phases of the observed
target, $\varphi$ a 8-component vector representing the phase in the
pupil (one aperture is chosen as a reference), and $\mathbf{A}$ is a
36 $\times$ 8 transfer matrix, whose properties form the core of this
discussion.

For this example, each row of $\mathbf{A}$ is essentially filled with
zeros, except for two positions corresponding to the apertures
making the baseline, that respectively contain $+1$ and $-1$.
In general, a closure relation is a linear combination of rows of
$\mathbf{A}$ that give the zero vector. The closure-phase is a special
case of one such linear relation, that simply adds together the
appropriate series of three rows to give the zero vector.
More complex relations involving more than three rows of $\mathbf{A}$
can however be produced. 

The total number of independent relations however remains constant,
and is exactly 28 \footnote{$(N-1) \times (N-2) / 2$, where $N$ is the
  number of sub-apertures in the pupil in this NRM case} in this
scenario. Using some algebra jargon is not inappropriate here: these
closure relations form a basis for the left-hand null space (or
Kernel) of $\mathbf{A}$. These relations can be summarized by a left
hand operator $\mathbf{K}$ that acts on $\mathbf{A}$, so that:

\begin{equation}
  \mathbf{K} \cdot \mathbf{A} = \mathbf{0}.
  \label{eq:K}
\end{equation}

Closure phase is convenient and easy to grasp. Moreover, it is a
natural choice and the only possible closure relation when the pupil
is made of only three sub-apertures. In practice for a baseline-rich
pupil like the 9-hole case used as an example, closure-phase alone
is not the best solution, as similar triangles in the pupil do exhibit
correlated closure-phases.
In addition to the closure phases, for NRM-interferometry data
reduction, it is also customary to carry a correlation matrix, that is
used during the modeling stage to produce reliable constraints on the
parameters of the fit. Using this idea of generalized closure-phase,
it is possible to find directly the relations that produce
decorrelated observables. The next section will show how to do this
for an arbitrary pupil shape.

\section{Kernel-phase}
\label{sec:kphi}

\begin{figure}
\centerline{\includegraphics[width=0.9\textwidth]
  {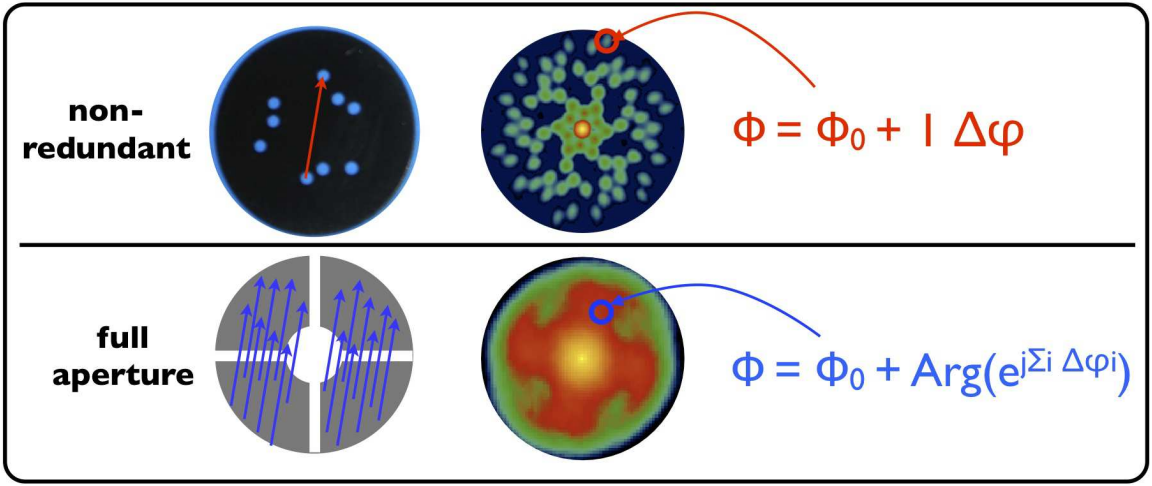}}
\caption{Comparison of the uv-plane phase measured with a NRM mask and
  with a conventional telescope aperture. In the latter scenario, a
  given baseline is sampled a large number of times within the
  pupil. The resulting expression for the phase is much more complex
  than in the non-redundant case: solving the deconvolution problem
  mentioned in the introduction, would mean being able to separate
  $\Phi_0$ from a sum of random phasors. While this problem is indeed
  degenerate, AO makes it possible to extract closure-phase like
  quantities that do not depend on wavefront residuals (see text).
}
\label{fig:red}
\end{figure}

For arbitrarily shaped pupils (see the comparison NRM vs full aperture
in Fig. \ref{fig:red}), there is an extra complication, as the useful
interferometric signal $\Phi_0$ finds itself buried under multiple
phase error contributions, which results into a fairly complex
(non-linear) expression for the uv-phase:

\begin{equation}
  \Phi^k = \Phi_0^k + \mathrm{Arg}(e^{j\Sigma_i \Delta\varphi_i}),
  \label{eq:arg}
\end{equation}

\noindent
where $i$ is an index to keep track of the $r$ identical baselines in
the pupil, contributing to the same region of the uv plane.
With a good wavefront correction (this approach has been
validated both on space-based and ground-based data) -
eq. \ref{eq:arg} can be linearized as follows:

\begin{equation}
  \Phi^k = \Phi_0^k + (1/r) \Sigma_i \Delta\varphi_i,
  \label{eq:lin}
\end{equation}

\noindent
and the entire problem can again be written in the matrix form of
eq. \ref{eq:transfer}, with a modified transfer matrix. In order to
maintain some resemblance with the previous case, it has been prefered
so far to split this transfer matrix into a product of two matrices:
a diagonal matrix $1/\mathbf{R}$ that encodes the redundance of the
baselines, and $\mathbf{A}$. In the (redundant) full aperture case,
each row of $\mathbf{A}$ now contains more than just two non-zero
values, that are however still either $-1$ or $1$. The problem is now
written as follows:

\begin{equation}
  \mathbf{R} \cdot \Phi = \mathbf{A} \cdot \varphi + \mathbf{R} \cdot
  \Phi_0.
  \label{eq:new_pblm}
\end{equation}

Even with this new problem, it is possible to find a left hand
operator $\mathbf{K}$ that verifies eq. \ref{eq:K}. While it is
possible to find by hand friendly closure-phase like relations if the
pupil geometry is not too complex, it quickly becomes difficult as the
matrix $\mathbf{A}$ can get quite big.
A very efficient way to achieve this goal is to calculate the singular
value decomposition (SVD) of $\mathbf{A}^T$. If the phase in the pupil
can be accurately described by a vector of size $N-1$, that produces
$M$ distinct uv points (some of which are highly redundant), the SVD
algorithm \cite{2002nreb.book.....V} allows to decompose the
$(N-1)\times M$ matrix $\mathbf{A}^T$ as the product of a $(N-1)\times
M$ column-orthogonal matrix $\mathbf{U}$, an $M \times M$ diagonal
matrix $\mathbf{W}$ with positibe or zero elements (the so-called
singular values), and the transpose of an $M \times M$ orthogonal
matrix $V$:

\begin{equation}
  \mathbf{A}^T = \mathbf{U} \cdot \mathbf{W} \cdot \mathbf{V}^T.
\end{equation}

Among other properties, the SVD explicitly constructs an orthonormal
basis for the null-space of a matrix: with the notations used in this
paper, the columns of $\mathbf{V}$ that correspond to singular values
equal to zero are exactly what needs to fill in the rows of
$\mathbf{K}$. These relations, conveniently orthogonal, when applied
to Eq. \ref{eq:new_pblm} produce observable quantities that just like
closure phase, are independent from instrumental and atmospheric
phase. These observable quantities are called Kernel-phases, since
they relate to the Kernel of the transfer matrix $\mathbf{A}$. The
following Section will present results of application of this
method.

\section{First ground-based results using Kernel-phase}

\begin{figure}
\centerline{\includegraphics[width=0.9\textwidth]
  {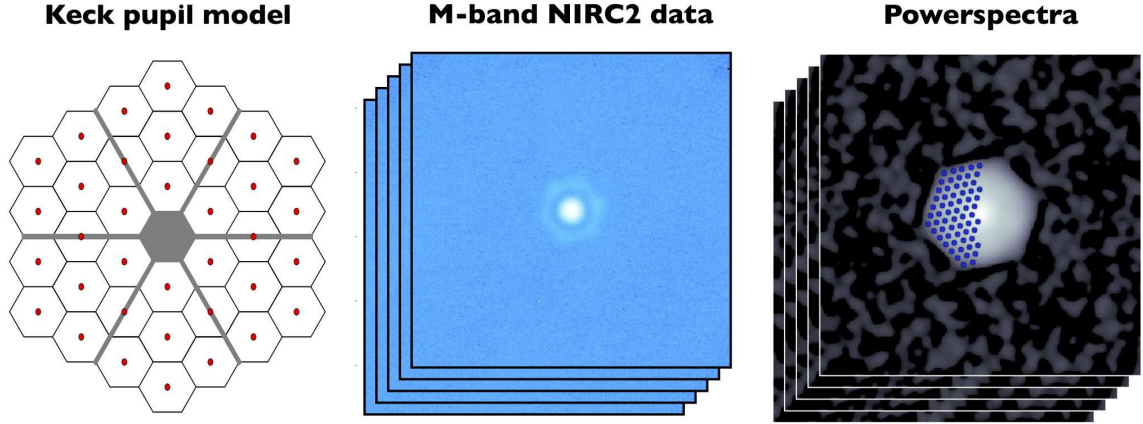}}
\caption{Left panel: Model of the Keck II Telescope pupil used for the
  determination of Kernel-phase relations. Each of the 36 segments of
  the pupil is reduced to a single point.
  Middle panel: example of M-band frame acquired with NIRC2.
  Right panel: each frame is Fourier-transformed, and this panel shows
  the powerspectrum (square modulus of the complex function), in
  non-linear scale. The uv sample points matching the model of the
  pupil are overlaid. For these locations of the uv-plane, the phase
  is extracted and combined into Kernel-phases using the relations
  introduced in Section \ref{sec:kphi}.
}
\label{fig:data}
\end{figure}

The Kernel-phase approach was successfully applied to two series of
narrow-band images acquired with the Near Infrared Camera and
Multiobject Spectrometer (NICMOS) onboard the Hubble Space Telescope
\cite{2010ApJ...724..464M}. In one case, a positive detection of a
known $10:1$ binary companion (GJ 164 B) with milli-arcsecond
astrometric prediction, at angular separation 0.6 $\lambda/D$ was
made.
In the other case, supposed to be a single calibration star (SAO
179809), a series of Monte Carlo simulations demonstrated that at
angular separation 0.5 $\lambda/D$ (i.e. 80 mas at $\lambda = 1.9
\mu$m), a contrast limit better than $50:1$ is possible at the 99 \%
confidence level. At angular separation 1 $\lambda/D$, this limit
increases to $200:1$ at the same confidence level.
For this paper, the technique was applied to a series of M-band images
acquired with NIRC2 on the Keck II telescope of the W.M. Keck
Observatory.

Application of the Kernel-phase technique starts with building a model
of the pupil of the telescope. The left panel of Fig. \ref{fig:data}
shows the model used for this data analysis: the segmented aperture of
the Keck II Telescope is reduced to a series of 36 points. The pupil
phase $\varphi$ is therefore assembled into a 35-component vector.
This 35-component pupil phase vector maps in the uv plane onto a
hexagonal grid of 63 distinct points, shown in the right panel of
Fig. \ref{fig:data}.
The transfer matrix $\mathbf{A}$ is therefore a $35 \times 63$
rectangular matrix. The SVD reveals that 18 singular values are
non-zero, leaving $63-18=45$ Kernel-phase relations.

Once the uv-points chosen and the matching Kernel-phase relations
identified, they can be saved in a template and used after extraction
of the phase from the data. The data used in this example consist of
one target and two calibrators\footnote{data courtesy of Adam Kraus
  and Michael Ireland}.
For each source, there is an average of 20 frames. After dark
subtraction, and flat-fielding, each frame is simply Fourier
transformed. The left hand side of the equation \ref{eq:new_pblm}
(i.e. $\mathbf{R} \cdot \Phi$) can be acquired by reading directly
the imaginary part of the complex visibility (the Fourier transform),
at the pre-defined uv-plane locations. The uncertainty associated with
the measurement of each sample is estimated from the dispersion of the
signal in the direct neighborhood of the uv-plane sample points.

Kernel-phases are constructed using the pre-determined relations for
each frame, then averaged per source, and uncertainties are
propagated. The final retained series of 45 Kernel-phases is the
weighted average for all frames. Just like for closure-phase,
calibration of Kernel-phase is simply a matter of subtracting the
Kernel-phase of one or more calibrators from the ones recorded on the
source of interest.

\begin{figure}
\centerline{\includegraphics[width=0.9\textwidth]
  {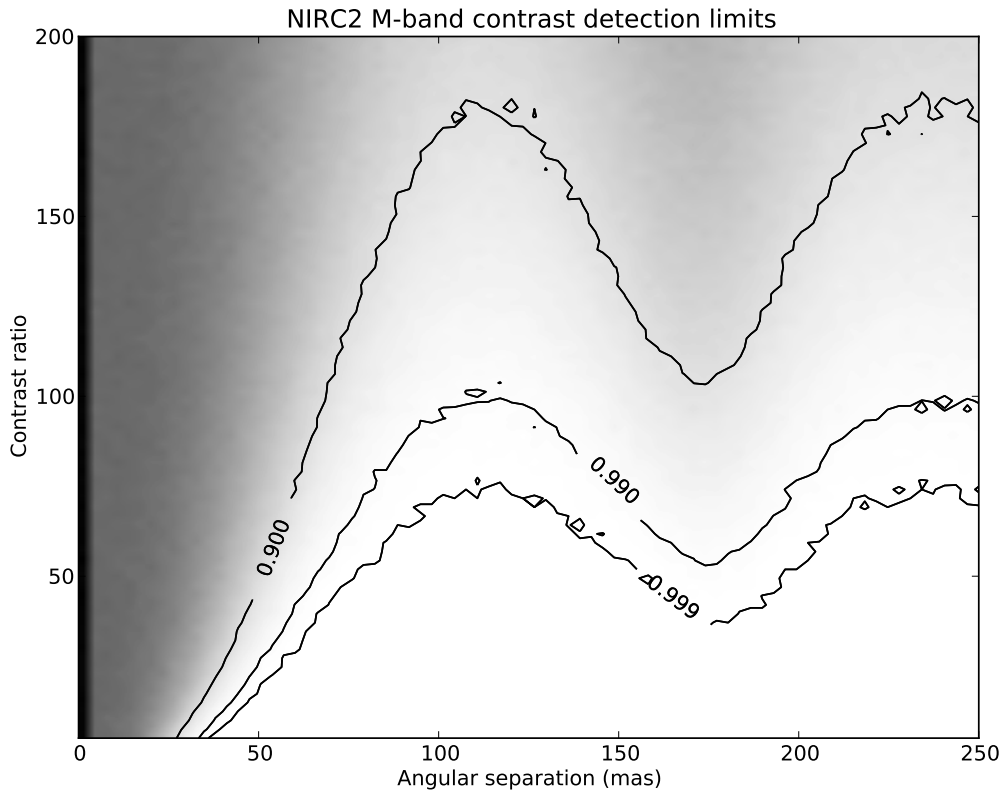}}
\caption{Levels of confidence in the detection of a companion in the
  (angular separation - contrast) space, using the proposed Kernel-phase
  algorithm on M-band data acquired at the Keck II Telescope, using NIRC2.
  A darker color indicates a region of lower confidence level. Three
  levels are highlighted: the 90 \%, 99 \% and 99.9 \% confidence
  levels. At angular separation 0.5 $\lambda/D$, a 20:1 contrast
  detection appears feasible at the 99.9 \% confidence level.
  Note that the highlighted contours closely match the radial
  evolution of the diffraction pattern by the Keck II pupil.
}
\label{fig:limits}
\end{figure}

While, no obvious detection can be reported from this data analysis,
the final calibrated closure phases were used as input in a Monte
Carlo simulation to determine contrast detection limits.

The sampling in the uv-plane (cf. Fig. \ref{fig:data}) is sufficiently
uniform for the sensitivity to be only a very weak function of the
position angle of a potential companion. Sensitivity is however
expected to be a strong function of angular separation: it is not
difficult to believe that the detection of a companion will be easier
if it is located between two diffraction rings than if it lands
exactly on a ring.
Note that the sampling also limits the outer working angle of the
analysis to $\sim 4 \lambda/D$, beyond which direct analysis of the
image is expected to provide superior results anyway.

Fig. \ref{fig:limits} summarizes the detection limits with this
data-set. The overall structure of this sensitivity 2D plot is
strikingly similar to what was reported after a comparable analysis of
NICMOS data \cite{2010ApJ...724..464M}.
The three highlighted confidence levels closely match the radial
evolution of the Airy pattern for Keck II. At 0.5 $\lambda/D$, a 20:1
contrast detection appears feasible with a high (99.9 \%) confidence
level. At 1 $\lambda/D$, this limit rises to 70:1.

While quite encouraging, it is not clear yet what is constraining the
detection limits on this dataset. In the near future, a comprehensive
set of simulations including background, readout and photon noise in
addition to wavefront errors will help define observing strategies
that will maximize the Kernel-phase's potential for high contrast
detections at the diffraction limit and in the super-resolution
regime.

\section{Conclusion}

Closure-phase was shown to be a special case of a wider family of
observable quantities immunte to phase noise and non-common
path errors. While invented for, and perfectly adapted to
interferometers with a limited number of apertures (with a minimum of
three, that is), closure-phase is not necessarily the most appropriate
of these observable quantities when dealing with baseline-rich
apertures of modern NRM-interferometry.
Using a very general linear algebraic approach, it was shown that the
notion of closure-phase can be generalized, and this generalization
extends the applicability of the technique to conventional
full-aperture images, if wavefront quality is sufficient.

Completing already published work that demonstrated potential of
the method on archive narrow-band HST/NICMOS data, and achieved
moderate contrast detection in the super resolution regime with good
astrometric precision, 
this paper showed that Kernel-phases can now be extracted from ground
based broad-band high-Strehl AO data, opening the doors of the
super-resolution regime that has so far been the exclusivity of
NRM-interferometry, even with a highly redundant pupil.

In addition to the already available HST/NICMOS archive data awaiting
re-analysis with the potential to lead to new detections in the
super-resolution regime and/or to improved detection limits, there are
multiple types of ground-based AO observing programs, mostly in L- and
M-band where the AO correction is at its best, that can now also take
advantage of this technique.

\acknowledgements{The author thanks Adam Kraus and Michael Ireland for
sharing the NIRC2 data-set used for this demonstration of the
technique done in this paper. This work was supported in part by the
Jet Propulsion Laboratory uner contract 1379504.}

\bibliographystyle{spiebib}
\bibliography{ms}
\end{document}